\documentclass[conference]{IEEEtran}
\IEEEoverridecommandlockouts

\usepackage{cite}
\usepackage{amsmath,amssymb,amsfonts}
\usepackage{algorithmic}
\usepackage{graphicx}
\usepackage{textcomp}
\usepackage{xcolor}
\usepackage{comment}
\usepackage{xspace}
\def\BibTeX{{\rm B\kern-.05em{\sc i\kern-.025em b}\kern-.08em
    T\kern-.1667em\lower.7ex\hbox{E}\kern-.125emX}}

\usepackage{accessibility}

\usepackage{enumitem}
\usepackage{tabularx}
\usepackage{booktabs}
\usepackage{float} 
\usepackage[numbers]{natbib}
\newcommand{\rebuttal}[1]{#1}

\newcommand{\rqone}{What is the hackathon's participant profile?}
\newcommand{\rqtwo}{What are the participants' motivations and expectations in participating in an affirmative hackathon?}
\newcommand{\rqthree}{What are the main challenges participants faced during the hackathon?}
\newcommand{\rqfour} {What are the key factors that influenced in participants' successful collaboration and satisfaction?}

\newcommand{\companyName}{Zup Innovation\xspace}

\begin{document}

\title{Affirmative Hackathon for Software Developers with Disabilities: An Industry Initiative}

\author{\IEEEauthorblockN{Thayssa Rocha}
\IEEEauthorblockA{\textit{Zup Innovation \& UFPA} \\
Belém, PA \\
thayssa.rocha@zup.com.br}
\and
\IEEEauthorblockN{Nicole Davila}
\IEEEauthorblockA{\textit{Zup Innovation \& UFRGS} \\
Porto Alegre, RS \\
nicole.davila@zup.com.br}
\and
\IEEEauthorblockN{Rafaella Vaccari}
\IEEEauthorblockA{\textit{Zup Innovation} \\
São Paulo, SP \\
rafaella.vaccari@zup.com.br}
\and
\IEEEauthorblockN{Nicoly Menezes}
\IEEEauthorblockA{\textit{UFPA} \\
Belém, PA \\
nicoly.menezes@icen.ufpa.br}
\and
\IEEEauthorblockN{Marcelle Mota}
\IEEEauthorblockA{\textit{UFPA} \\
Belém, PA \\
mpmota@ufpa.br}
\and
\IEEEauthorblockN{Edward Monteiro}
\IEEEauthorblockA{\textit{StackSpot} \\
São Paulo, SP \\
edward.monteiro@stackspot.com}
\and
\IEEEauthorblockN{Cleidson R. B. de Souza}
\IEEEauthorblockA{\textit{UFPA} \\
Belém, PA \\
cleidson.desouza@acm.org}
\and
\IEEEauthorblockN{Gustavo Pinto}
\IEEEauthorblockA{\textit{Zup Innovation \& UFPA} \\
Belém, PA \\
gustavo.pinto@zup.com.br}
}

\maketitle


\begin{abstract}
People with disabilities (PWD) often encounter several barriers to becoming employed. A growing body of evidence in software development highlights the benefits of diversity and inclusion in the field. However, recruiting, hiring, and fostering a supportive environment for PWD remains challenging. These challenges are exacerbated by the lack of skilled professionals with experience in inclusive hiring and management, which prevents companies from effectively increasing PWD representation on software development teams. 
Inspired by the strategy adopted in some technology companies that attract talent through hackathons and training courses, this paper reports the experience of Zup Innovation, a Brazilian software company, in hosting a fully remote affirmative hackathon with 50 participants to attract PWD developers. This event resulted in 10 new hires and 146 people added to the company's talent pool.
Through surveys with participants, we gathered attendees' perceptions and experiences, aiming to improve future hackathons and similar initiatives by providing insights on accessibility and collaboration. Our findings offer lessons for other companies seeking to address similar challenges and promote greater inclusion in tech teams.

\end{abstract}


\section{Introduction}

People with disabilities (PWD) often face several barriers in their lives. According to the World Health Organization (WHO)~\cite{world2011world}, lack of accessibility, inadequate policies and standards that take into account the needs of PWDs, and lack of evidence to understand this context are some of the several difficulties faced by this group. There is also evidence of disparities in the workplace and barriers PWD face to becoming employed, such as lower job security and less job flexibility~\cite{schur2017disability}. Addressing these barriers is relevant to promoting equity, diversity, and inclusion for this group in society~\cite{PractingEDI}. As for companies, the benefits of hiring PWD include improvements in profitability, competitive advantage, and an inclusive work culture~\cite{lindsay2018systematic}. 

In recent years, the technology sector has emerged as a dynamic field with opportunities for innovation, including solutions to support PWD. However, this group remains underrepresented in software development, with some authors suggesting a diversity~\cite{albusays2021diversity} and inclusion~\cite{rocha2024effective} crises. The tech industry's fast-paced, constantly evolving nature presents unique challenges for PWD, who may require additional support or accommodations to thrive~\cite{cardoso2023supporting}. Workarounds have been observed for the effective inclusion of PWDs in software development teams, such as adapting tools and adjusting work dynamics, including the need to ask for help from coworkers, whether they have disabilities or not~\cite{rocha2024effective}. Despite these efforts and the technology potential to create accessibility improvements to PWD, the gap between opportunity and inclusion in the tech sector still persists.

The problem of helping PWD to join the tech workforce is a two-sided coin. On the one hand, for PWD, finding employment in the technology sector is often fraught with difficulty. The lack of inclusive hiring practices, insufficient accommodations during recruitment processes, and persistent stereotypes regarding their capabilities are barriers often perceived~\cite{A379, A520}. On the other hand, many tech companies struggle to implement the necessary steps to recruit and retain PWD talent effectively. For instance, some studies~\cite{A543} have shown that there is an inadequate outreach effort and minimal awareness of how to engage this talent pool. 

In this work, we report on our experience designing and applying an online tech \textit{Affirmative Hackathon for PWD} organized by \companyName, a large Brazilian tech company. During the ten-day event,  we collected participant data through a series of surveys, to ensure the results were connected to the company's 
two main goals: (1) to empower PWD by enhancing their 
knowledge
and skill sets as demanded by modern software development companies, and (2) to enrich the company's pool of talented PWD candidates, thereby strengthening the pipeline for future recruitment processes. By focusing on these goals, this hackathon sought to provide valuable learning experiences for the participants and address the industry's need for greater inclusion and representation in the tech workforce.

Based on the analysis of the collected data, this
paper makes the following contributions:
\begin{itemize}
    \item \textbf{Empirical results} from a real experience of a corporate online hackathon dedicated to PWD; and
    \item  An opportunity to \textbf{discuss the importance of affirmative actions to promote diversity and inclusion} in software development teams.
\end{itemize}

Our results suggest that applying inclusive measures to support participants with disabilities in an exclusive online hackathon can promote a 
learning environment and foster networking and employability to this public.

\section{Related Work}

To understand how corporate hackathons are organized in technology companies, Valença and colleagues~\cite{9226320} conducted a systematic mapping focused on identifying the hackathon process involved in the stages before and after the events, as well as the main characteristics of corporate hackathons. \rebuttal{Furthermore, this work is limited by the lack of empirical analyses 
of individuals directly involved in these events, which would deepen the understanding of the dynamics involved.}


Paganini and Gama~\cite{kievHackathons} explored the reasons behind the low representation of women in hackathons, analyzing data from a hackathon organized by female undergraduate students to female students,  in order to identify gender barriers and motivations affecting women's participation. Challenges that emerged included low self-esteem, insecurity about technical skills, and the overwhelming presence of males at such events. In addition, women were motivated to participate by opportunities to learn new skills, build networks, and gain practical experience. By creating a welcoming, safe space with female mentorship and engaging activities, participants reported feeling more comfortable and encouraged. 

To explore the inclusion and participation of minority groups in hackathons, Prado and colleagues~\cite{TransHackathon} conducted a study focused on transgender, non-binary, and non-conforming communities' experiences to identify the specific challenges these individuals face at such events. The study analyzed responses from 44 participants and included semi-structured interviews. Findings revealed issues such as discrimination, lack of representation, and the absence of effective policies to combat LGBTQIA+ phobia. To address these challenges, the authors recommended incorporating transgender and non-conforming individuals into organizing teams, adopting inclusive communication, and providing adequate support. 

\rebuttal{The earlier studies do not specifically investigate the experiences of PWD in hackathons. Therefore, this study aims to fill this gap by stimulating discussions about an empirical experience and inclusive measures designed to support this group in hackathons specifically tailored for them. 
}

\section{The Affirmative Hackathon}\label{sec:hackathon}

In this section, we discuss the planning and execution of the \textit{Affirmative Hackathon for People with Disabilities}, detailing the motivations behind its creation, participant selection criteria, event phases, and internal organization strategies implemented by \companyName.

\subsection{Motivation}

The company \companyName organized the \textit{Affirmative Hackathon for People with Disabilities} in the second semester of 2024, committed to promoting diversity and inclusion and building a talent pool for future recruitment processes. This initiative was part of the company’s efforts to meet its strategic goal of diverse teams, besides complying with the legal obligation by Brazilian law, which requires hiring a percentage of PWDs in companies with more than 100 employees~\cite{Leidecotas}.

The \textit{Affirmative Hackathon} was presented as an opportunity to foster a more inclusive work environment by bringing different perspectives and skills to the company. Diversity, especially in the technology sector, is essential for driving innovation and solving problems more creatively and efficiently~\cite{PractingEDI, PrickladnickDiversity}. The event aimed to empower and engage people with disabilities in challenging projects, allowing them to demonstrate their technical skills and potential and connecting them to \companyName's organizational culture. Besides, employers could volunteer as mentors or instructors in the initiative, increasing their awareness of accessibility and inclusion. Finally, by building a PWD talent pool, the company prepared for future recruitment processes, aiming to form more inclusive and diverse teams, which is crucial for the success and sustainable growth of the organization.

\subsection{Hackathon Preparation Phase}\label{sec:preparation}
The event was planned to be a ten-day problem-solving marathon with collaborative and intensive characteristics. The minimum requirements for participants enrollment were:
\begin{itemize}
    \item Be a person with a disability, aged 18 or older, and a resident of Brazil;
    \item Knowledge in Kotlin/Java or FullStack, and tools such as Git and Docker, whether or not they have professional experience;
    \item Availability to attend the online sessions, collaborate with a team to develop the challenge solution, and present it to the evaluation panel; 
    \item Access to a computer with a stable Internet connection throughout the selection process and the duration of the hackathon; and
    \item Is not currently working at the company.
\end{itemize}

The hackathon was publicized through the company's official LinkedIn and Instagram. The company also developed a campaign encouraging employees to publicize it on their social networks. Professionals and Non-governmental organizations (NGOs) who are references in Brazil for publishing PWD content in technology were also contacted and asked to repost information about the hackathon on their social networks. In total, were contacted 79 NGOs focused on the PWD public, such as IBC (Instituto Benjamin Constant), a Brazilian federal institution specialized in the education and care of blind and visually impaired people~\cite{govbr_ibc}.

We received 211 applications. Each application went through a two-step evaluation process, which is detailed next. 
 
\textbf{Step 1: Application Validation.} Here, the organization's committee evaluated the minimum knowledge and PWD information required. Among those 211 applicants, 132 were eliminated due to not meeting the enrollment criteria for the event. In particular, four applicants were already company employees, 48 applicants did not declare themselves to be disabled, and 80 participants did not have the minimum technical knowledge defined as necessary to participate. The remaining 79 applications were considered successful and were passed to the next step.
 
\textbf{Step 2: Knowledge Test.} This evaluation consisted of an online theoretical knowledge test to ensure they met the minimum technical criteria for developing the hackathon's challenges. This test included 60 questions about software engineering best practices and technologies required from the hackathon participants. Of the 60 questions, 16 were applied to all candidates, and the remaining questions were applied based on the professional experience reported on the application form. For instance, clean code questions were presented to all participants, while questions on the usage of Spring Boot in Java projects were only shown to those senior developers. From the 79 preapproved in Step 1, 50 were selected for the event. The remaining 29 were not considered either due to missing the knowledge test or insufficient performance on it.

\subsection{Hackathon Phase}

Starting on \textit{Day One}, participants were divided into ten groups of five people, and each group received one challenge consisting of a problem that should be solved by designing a software solution. Examples of challenges included the production of accessible software documentation using Artificial Intelligence (AI) and using AI to assist in the correction and analysis of written test answers. A tech-skilled employer was assigned as a mentor for each group to support them in achieving their goal. During the ten days, the program offered two hours of synchronous online sessions, known as \textit{knowledge pills}. In these sessions, the company promoted knowledge-sharing presentations intended to provide content that could assist the groups in solving the challenges.

\begin{figure}[h!]
    \centering
    \includegraphics[width=1\linewidth, alt={The image is a flowchart illustrating the phases and steps of a hackathon process. It is divided into two main phases: the Preparation phase and the Hackathon phase. In the Preparation phase there are three steps: Publicity and registration: 211 applications are received. Application Validation: 79 valid applications are identified. Knowledge Test: 50 approved applications are selected. In the Hackathon phase there are also three steps: Day One: The hackathon begins. Hackathon Activities, that include: Online knowledge pills sessions (2 hours per day), Self-organized group meetings, Challenge development activities and Mentoring support. Demo Day: The hackathon concludes with a demonstration day.}]{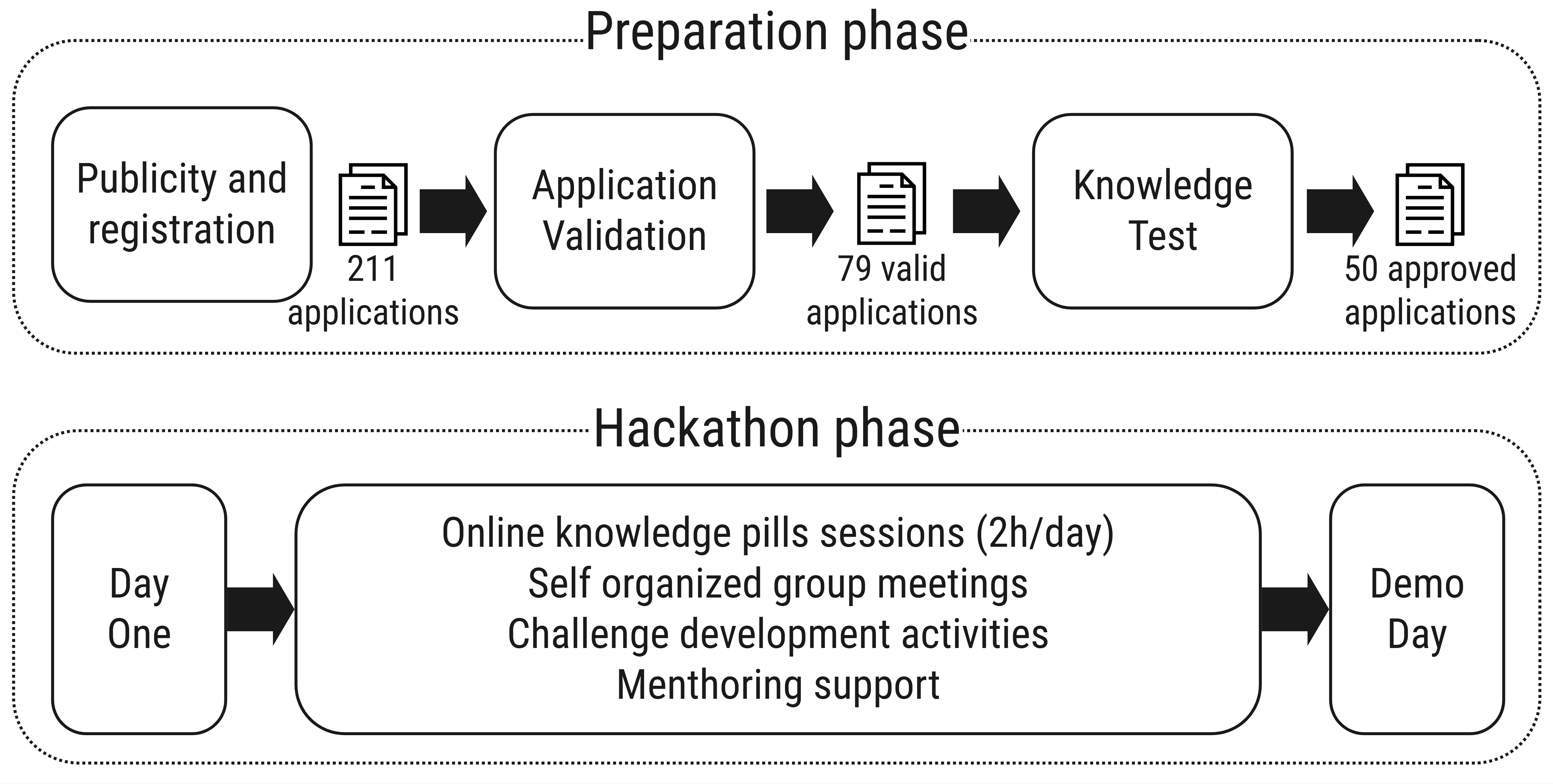}
    \caption{Hackathon's phases.}
    \label{fig:phases}
\end{figure}

Similarly to other hackathons~\cite{PeThan2019}, the event culminated in a \textit{Demo Day}, where the teams presented their solutions to a judging panel of tech-skilled employers. The event and pre-event phases are described in Figure ~\ref{fig:phases}.


\companyName guaranteed accessibility resources, such as breaks during online sessions, sign language interpreters, and screen reader compatibility. As the hackathon occurred remotely, the official video conference tool -- Google Meet -- was also chosen based on accessibility compliance. Prizes were offered to the group members of the three best-evaluated software solutions, besides certificates to all participants who presented their solutions to the judging panel. 

\subsection{Internal Hackathon Organization}
The \companyName ``diversity and inclusion'' internal department coordinated the hackathon. This department counted on the support of marketing, data and analytics, legal and compliance areas. Some employees expressed interest in joining the initiative to work as volunteers. In total, 31 \companyName employees collaborated in the event. In particular:

\begin{itemize}
    \item One general coordinator of the initiative;
    \item Five employees from the diversity and inclusion team to support the participants and manage communications;
    \item 25 volunteers: 13 as instructors who facilitated the \textit{knowledge pills} online sessions and 12 as mentors, who helped the teams on building the challenge's solution.
\end{itemize}

\section{Method}
\label{sec:method}
In this section we present the details of the method used to conduct the research.
\subsection{Research Questions}\label{sec:method}

The research questions aim to guide our ultimate goal of identifying
the factors that should be considered when promoting 
affirmative hackathons to guarantee participant satisfaction and the achievement of the company's objectives. They are:

\begin{itemize} [leftmargin=0.9cm]
   \item[RQ1] \rqone
   \item[RQ2] \rqtwo 
   \item[RQ3] \rqthree
   \item[RQ4] \rqfour
\end{itemize}

Each question addresses core aspects of the participants' experience and event organization. Understanding the participant profile (RQ1) is essential for assessing whether this type of hackathon attracts the intended audience and identifying their specific needs. Examining participants' motivations and expectations (RQ2) supports alignment between event activities and participants' professional growth goals. 
Evaluating participants' main challenges (RQ3) helps understand and design more effective and inclusive environments. Finally, identifying factors that foster successful collaboration and participant satisfaction (RQ4) aids in refining strategies to enhance teamwork, learning, and recruiting 
PWD.

\subsection{Family of Surveys}
In this study, we employed a structured series of surveys to guide the participant selection process and assess their experiences throughout the event. The first survey (Section~\ref{sec:survey1}) was designed to initiate the selection process by collecting key demographic information from the target population. This data helped us promote a diverse and representative group by ensuring that only PWD with knowledge of the target technologies would participate in the hackathon.

Following the participant selection, on \textit{Day One}, we administered a second survey (Section~\ref{sec:survey2}) to delve deeper into the participants' backgrounds, motivations, and expectations for the \textit{Affirmative Hackathon}. This allowed us to tailor the event to address their needs and aspirations, also creating a baseline for the final evaluation.

Lastly, on \textit{Demo Day} we conducted a final survey (Section~\ref{sec:survey3}) to evaluate the participants' overall experience at the end of the hackathon. This survey gathered feedback on various aspects, including the structure and dynamics of the activities, the quality of the infrastructure, accessibility measures, mentor support, group communication, and participants' satisfaction. \rebuttal{All surveys were developed in Portuguese, and their results were subsequently translated into English.}

\begin{figure}[h]
    \centering
    \includegraphics[width=1\linewidth, alt={The image is a flowchart depicting the process and phases of a hackathon event, divided into the Preparation Phase and Hackathon Phase. Preparation Phase: Starts with 211 applications and after Application Validation, 50 applications are approved. Survey 1 generated with 50 answers. Hackathon Phase: Begins with Day One having 50 participants and survey 2 conducted with 24 anonymous answers. Followed the Day one goes 10 days of Sync and Async Hackathon Activities. After that, the Hackathon ends with Demo Day, with 31 participants where Survey 3 is conducted and collected 27 answers also anonymous. This image outlines the structured process of a hackathon, showing the progression from application to participation and feedback collection through surveys.}]{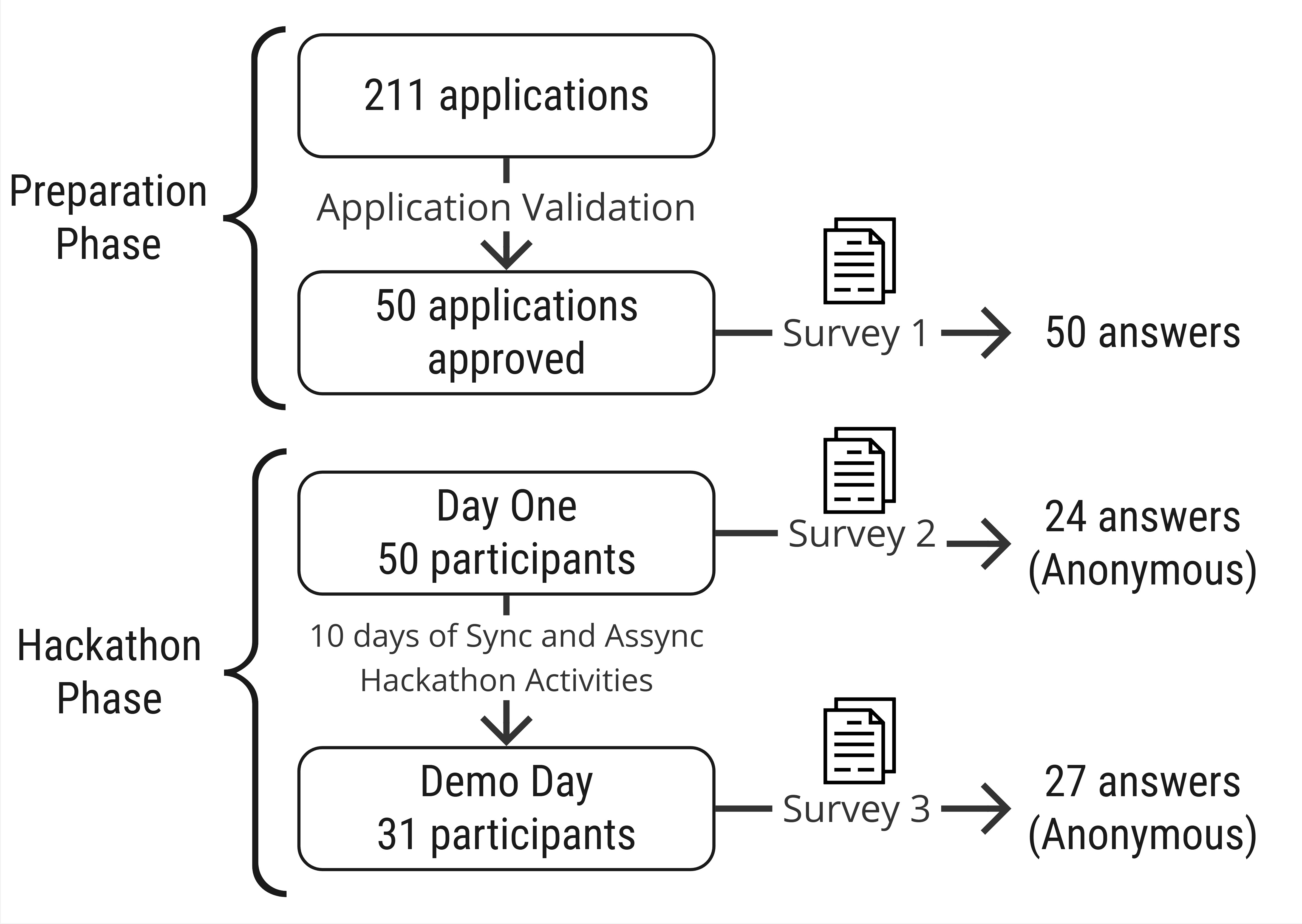}
    \caption{Surveys and answers.}
    \label{fig:surveys}
\end{figure}

As previously described, the surveys were applied in different stages of the hackathon. They also had a different number of answers (Figure~\ref{fig:surveys}).
It is important to note that \textit{Survey 1} was the only one we kept participant's identification since it was connected to the hackathon application's process. However, data for this research analysis was anonymized. Regarding \textit{Survey 2} and \textit{Survey 3}, data collected was already anonymous, and answering the surveys was optional. \rebuttal{For this reason, in Section \ref{sec:rq2}, \ref{sec:rq3}, and \ref{sec:rq4} when we present citations, the participant id is indexed by each survey. As a result, we identified participants as S2P1 to S2P24 for \textit{Survey 2} and S3P1 to S3P27 for \textit{Survey 3}.}

\subsubsection{Survey 1}\label{sec:survey1}

As the first questionnaire was intended to open the selection process for the hackathon, we asked demographic questions and more information about their experience in the tech industry \rebuttal{as well as information about accessibility issues, shown in Table~\ref{tbl:survey1-questions}}. The full questionnaire had 23 questions \rebuttal{and can be found in our supplemental material\footnote{https://anonymous.4open.science/r/HackathonPWD-31C4/}.}
During 38 days, the questionnaire received 211 responses.

\begin{table}[h]
\caption{Survey 1 - Accessibility questions\rebuttal{(none required)}.}

\label{tbl:survey1-questions}
\centering
\begin{tabularx}{\linewidth}{lXc}
\toprule
\textbf{ID} & \textbf{Question} & \textbf{\rebuttal{Possible values}}
\\ \midrule




S1Q18 & Do you need any accessibility support?  & \rebuttal{ Yes / }No \\
S1Q19 & Please specify the kind of accessibility support you need. & \rebuttal{Open-ended} \\
S1Q20 & Do you use any assistive technology to interact with digital environments? & \rebuttal{Yes / }No \\
S1Q21 & Please specify the assistive technology you use.& \rebuttal{Open-ended} \\

    
\bottomrule
\end{tabularx}
\end{table}

\subsubsection{Survey 2}\label{sec:survey2}

Our second survey was designed to gather preliminary insights from the hackathon participants (see Table~\ref{tbl:survey2-questions}). By understanding past experiences, motivations, and expectations, we aimed to enhance the overall experience during the event. Additionally, we wanted to identify potential challenges in group dynamics to mitigate eventual accessibility and inclusion challenges in our hackathon. 

Except for the first one, the questions in this survey were open-ended. They were analyzed using the techniques described in Section~\ref{sec:dataanalysis}. Based on the results of this analysis, we designed some questions for the next survey.

\begin{table}[h]
\caption{Survey 2 questions \rebuttal{(all required)}.}

\label{tbl:survey2-questions}
\centering
\begin{tabularx}{\linewidth}{lXc}
\toprule
\textbf{ID} & \textbf{Question} & \textbf{\rebuttal{Possible values}}
\\ \midrule
S2Q1 & Have you participated in other hackathons before? & Yes \rebuttal{/ No}\\
S2Q2 & What motivated you to sign up for this hackathon? & \rebuttal{Open-ended} \\
S2Q3 & What do you expect to experience in the coming days during our meetings and activities? & \rebuttal{Open-ended} \\
S2Q4 & What kinds of difficulties or challenges have you faced in group activities that you would like to avoid during this hackathon? & \rebuttal{Open-ended} \\
\bottomrule
\end{tabularx}
\end{table}

\subsubsection{Survey 3}\label{sec:survey3}

The third and final survey collected data at the end of the \textit{Demo Day}. 
It focused on assessing participants' experience at the event and gathering feedback on different \rebuttal{aspects through three satisfaction levels (Excellent / Good / Could be better). The evaluated aspects were: ``\textit{Knowledge pills}'' dynamics, event management, and technical mentors' performance.}
\rebuttal{Other questions were also posed to investigate participants' experience during the Hackathon. Table~\ref{tbl:survey3-questions} presents a set of questions (S3Q17 to S3Q24)}
based on feedback collected by \textit{Survey 2} in \textit{Day One}, i.e., regarding the main challenges participants faced in \textit{previous} collaborative experiences. In this group, the questions were presented using a 3-point Likert scale. \rebuttal{The full \textit{Survey 3} instrument is also available in the supplemental material}.

\begin{table}[h!]
\caption{Survey 3 questions \rebuttal{(all required)}.}
\label{tbl:survey3-questions}
\centering
\begin{tabularx}{\linewidth}{lX}
\toprule
\textbf{ID} & \textbf{Question} 
\\ \midrule
 
 



 S3Q17 & Communication with \companyName members\\
 S3Q18 & Communication between your group members\\
 S3Q19 & Empathy from \companyName members\\
 S3Q20 & Empathy between your group members\\
 S3Q21 & Organization by \companyName\\
 S3Q22 & Organization between your group members\\
 S3Q23 & Openness for questions and contributions with \companyName members\\
 S3Q24 & Openness for questions and contributions between your group members\\
 \bottomrule
\end{tabularx}
\end{table}

\subsection{\rebuttal{Surveys Elaboration, Validation and Authors involvement}}
\rebuttal{\textit{Survey 1} was built by \companyName ``diversity and inclusion'' department and validated internally by the general coordinator of the initiative, who is also one of this paper's authors. Adjustments in the question's order and answer options were made to provide the complete information needed by the company to manage the selection process. \textit{Surveys 2 and 3} were elaborated by the first author and validated by the ``diversity and inclusion'' department and the general coordinator of the initiative. No changes were requested for these instruments. Although five employees of Anonimous Tech are among the paper's authors, only two participated directly in the hackathon initiative: one as a "knowledge pill" facilitator and the other as the general coordinator.} 

\subsection{Data Analysis}
\label{sec:dataanalysis}
As mentioned in the previous sections, our surveys had closed and open questions. The \textit{closed} questions were analyzed by computing the descriptive statistics for the questions. As for the \textit{open} questions, we conducted an inductive qualitative analysis~\cite{creswell2013research}, coding the survey answers, aiming to understand the main topics associated with each question and, whenever possible, themes associated with different questions. The coding process was carried out \rebuttal{manually}
by the first author and subsequently reviewed and verified by the two last authors to ensure accuracy and consistency. The present codes are the result of this analysis.

\subsection{\rebuttal{Ethics}}
\rebuttal{When subscribing to the Hackathon (Survey 1) and answering the following surveys, participants were informed that the data provided would be used to research matters and to improve the initiative for subsequent editions. They were assured that no personal information would be used in a way that would make identification possible. The authors have a research project approved by the Research Ethics Committee of the UFPA that encompasses the company's authorization to perform research on this subject.}

\vspace{0.2cm}
\section{Results}
We group our results according to our research questions.

\subsection{RQ1: \rqone} 
In this research question, we provide an overview of the demographic, accessibility, and technical profiles of the 50 initial participants. It includes information on self-identified race, sexual orientation, and gender identity, as well as details about their disabilities and accessibility needs. Additionally, this section reports participants' levels of technical seniority, years of experience, and areas of technical focus. This information was gathered from Survey 1 (Section~\ref{sec:survey1}).

\vspace{0.2cm}
\noindent
\textbf{Race, Gender \& Sexual Orientation.} When asked about their race, participants self-identified as follows: 24 responded as White, 19 as Brown, five as Black, and one as Asian. One preferred not to answer. 
Regarding sexual orientation, 40 participants identified as heterosexual, four as bisexual, two as asexual, one as pansexual, and one as homosexual. Two participants did not answer this question. In terms of gender identity, 36 people identified as Cisgender\footnote{When a person identifies with the gender assigned at birth.} Man, six identified as Cisgender Woman, two identified as Non-binary\footnote{When a person does not identify as either man or woman.}, one as Transgender\footnote{When a person identifies differently from the gender assigned at birth.} Man, and one identified as other. Four people preferred not to answer about their gender identity.



\begin{figure}[h]
    \centering
    \includegraphics[width=1\linewidth, alt={This bar chart illustrates the number of individuals with various types of disabilities. The x-axis categories include: Physical, Neurodivergent, Visual, Hearing, Other, and Multiple. The y-axis displays the count of individuals, ranging from 0 to 20. The bars represent the following data: Physical (16 individuals), Neurodivergent (12 individuals), Visual (11 individuals), Hearing (10 individuals), Other (2 individuals), and Multiple (1 individual).}]{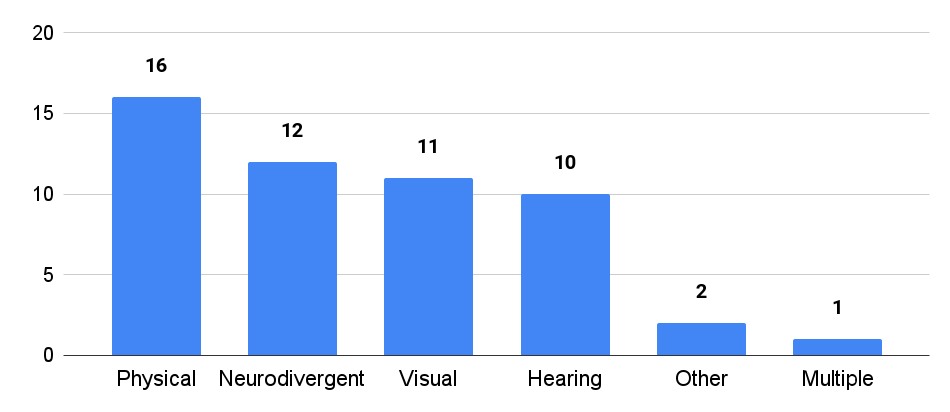}
    \caption{Disability distribution}
    \label{fig:disability}
\end{figure}

\vspace{0.2cm}
\noindent
\textbf{Disabilities \& Accessibility Resources.} 
Participants could select more than one disability, so the total amount of representation is 52 since two participants informed two types of disabilities each. They reported their disabilities as follows: Physical (16), Neurodiverse (12), Visual (11), Hearing (10), Other (2), and Multiple (1). Only three of the 50 initial participants do not have official documentation of their disability. Figure \ref{fig:disability} presents the distribution of participants based on their disabilities.

\noindent 
When asked about accessibility resources (S1Q18 and S1Q19), 36 participants responded, but only 13 stated that they needed additional resources. The participants highlighted some accessibility needs, emphasizing the presence of sign language interpreters, which nine people mentioned. Other necessities varied among requests for image descriptions and platforms accessible to screen readers. Additionally, the need for a wheelchair-accessible space, simple and objective written information about the topics discussed, to be emailed after the meeting, and considerations for adaptations for autistic individuals. In addition to these items, S1Q20 and S1Q21 inquired about the use of assistive technologies. Ten participants mentioned the use of assistive technologies, such as screen readers --- like NVDA ---, a notebook with an additional monitor and camera, magnifying glasses, and Zoom or Microsoft Teams for transcribing online meetings.

\begin{figure}[h!]
    \centering
    \includegraphics[width=1\linewidth,alt={The image is a bar chart displaying the different technologies related to the participant's technical profile. The x-axis lists the technologies, and the y-axis represents the number of participants, ranging from 0 to 50 and distributed as follows: Back-end Java (43), Fullstack (17), Back-end Kotlin (13), Back-end Python (14), Back-end Node.js(13), and Front-end React (13). The chart highlights highlighting with darker blue the three target technologies: Back-end Java, Fullstack and Back-end Kotlin.}]{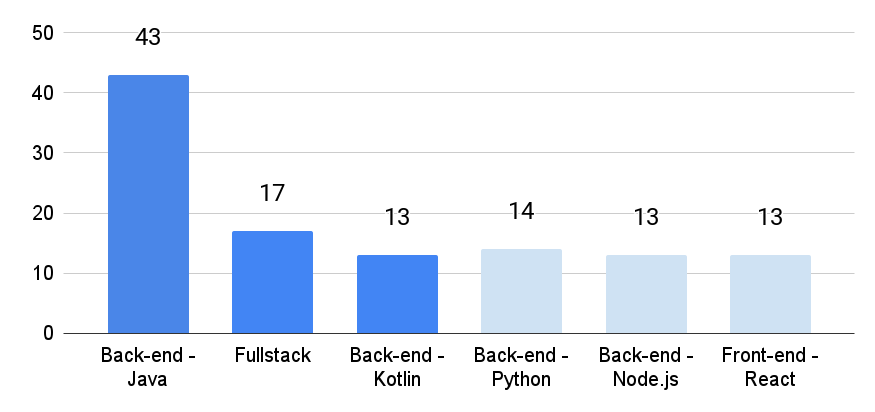}
    \caption{Technology distribution.}
    \label{fig:tech}
\end{figure}

\vspace{0.2cm}
\noindent
\textbf{Technical Profile.}  44 participants self-identified as Beginners, four as Mid-level, and two as Specialists. However, after the knowledge test results, the two Specialists were categorized as Seniors and two Mid-level were re-framed into beginners' profiles. 
Participants reported different levels of experience: 14 were classified with up to one year of experience; 20 participants were classified with more than one to five years of experience; four participants mentioned having more than five years of experience, while six mentioned only academic experience or no experience at all and finally, four did not fill this information.
The technical profile also varied significantly, especially among the three technologies targeted for this hackathon (Kotlin, Java, and Fullstack development). Almost 90\% of the participants declared knowledge as Back-end Java developers (43 of 50), followed by Fullstack (back and frontend technologies) with 17 occurrences and Back-end Kotlin with 13. Other technologies were also present in their technical profile declarations, being the most common Back-end Python with 14 participants, and Back-end Node.js and Front-end React both mentioned 13 times. Figure~\ref{fig:tech} presents this distribution, highlighting with darker blue the three target technologies. Some other technologies were less mentioned, such as: Quality Assurance (11), Cloud Computing (10), Data (9), DevOps/SRE (6), Artificial Intelligence/Machine Learning (4), and Information Security (3). 

\noindent
Among the 50 participants, only four had previously worked at \companyName, and 11 mentioned someone they know who is working now for the company. Regarding the experience with hackathons, from the 24 respondents in \textit{Survey 2}, only three had already participated in such initiatives. All others were experiencing a hackathon for the first time.

\vspace{0.2cm}
\noindent
\textbf{Participant's Profile.}
\label{sec:persona}
Based on the data obtained from the questionnaire, we designed a persona for the hackathon. Personas are fictitious characters used to represent users in terms of their goals and personal characteristics, and are commonly used by designers to turn data about people using software solutions more tangible~\cite{personasMarsden}. Here, we use it to characterize the most common profile of our hackathon's attendance.

\noindent
To create the persona, we analyzed data provided on participants' demographic characteristics, technical profiles, and accessibility needs and used the most common values. According to our analysis, a \textit{persona} for this hackathon would be: ``\textbf{White cisgender heterosexual man, with a physical disability, identified as a beginner software developer with one or two years of experience in Java technologies.}''

\subsection{RQ2: \rqtwo} 
\label{sec:rq2}

This section presents the themes we identified by the inductive qualitative analysis of the open questions on participants' motivations and expectations for joining the hackathon. In particular, we analyzed questions S2Q2 and S2Q3 from \textit{Survey 2} (see Table~\ref{tbl:survey2-questions}). We had 24 responses for this survey.

\vspace{0.2cm}
\noindent
\textbf{Motivations.} We identified four key reasons why participants decided to apply for this hackathon. They are described below.

Improve Technical Knowledge: 
Participants reported being motivated by the opportunity to learn and improve their technical knowledge, driven by a desire for personal and professional growth. For them, the \textit{Affirmative Hackathon} represents a way to enhance their skills, particularly in areas like accessibility and innovative technologies, such as AI, while also testing their abilities in challenging environments. \rebuttal{S2P15} reported \textit{``I received the email and thought it was a good opportunity to gain new knowledge. This was my first time using LLMS in software development.''}

Networking, Employment, and Career Development:
Participants described that joining a hackathon might be an interesting opportunity for career advancement, whether through networking, hands-on learning, or the possibility of joining companies like \companyName. For instance, when asked about their motivation, \rebuttal{S2P1} answered: \textit{``I identified an opportunity for possible career growth.''}

Challenge Themselves:
Participants reported the desire to challenge themselves by testing their technical abilities. They faced the hackathon as an opportunity to push their boundaries, test their skills, and gain valuable experience in real-world scenarios. Additionally, the sense of accomplishment from overcoming complex tasks and the potential for career advancement were significant driving factors. As \rebuttal{S2P1} reported: \textit{``What motivated me to sign up for this hackathon was my big interest in challenges, especially in programming and IT. I believe that technological innovation is constantly evolving and can always improve the performance of companies}.'' \rebuttal{S2P16} stated: \textit{``Learning to challenge myself is very good}.''

Engaging With Other PWD:
Participants mentioned that joining the
Affirmative Hackathon might promote the opportunity to collaborate with other PWD developers from different fields, as expressed by \rebuttal{S2P14}: ``\textit{Having this experience with other people who have disabilities, improving as a professional and having access to opportunities}.''


\vspace{0.2cm}
\noindent
\textbf{Expectations.} When asked about their expectations for the hackathon activities, participants shared different perspectives, sometimes very similar with some motivations, such as learning and the development of collaborative activities.

Build a Collaborative Learning Environment: Participants had expectations about the creation of a collaborative learning environment, centered on fostering personal and professional growth through shared experiences. This highlights the importance of dedicating time and patience to sharing complex concepts during the hackathon. The hope was that these interactions would enhance individual skills and create a supportive and cohesive team dynamic, exemplified by \rebuttal{S2P16} as: ``\textit{A mutually supportive environment where we can develop essential skills and strengthen our bonds as a team.}''

Learn From Others: There was a clear expectation about learning and gaining technical insights from both the sessions and the interactions with fellow participants. There was strong anticipation for demonstrations of various technologies and practical experiences to help solidify understanding. In summary, the participants were eager to absorb the concepts and tips shared by the presenters: ``\textit{I expect demonstrations of the technologies and some challenges}'' \rebuttal{(S2P20)}.

Learn Together: A commonly reported expectation is learning from other participants and exchanging experiences. These are centered around gaining technical knowledge through collaborative efforts and dynamic interactions. Respondents expected practical exercises, idea exchanges, and discussion engagement to foster mutual growth. In particular, given the focus on emerging techs, deepening their knowledge of such technologies was also highlighted by a few participants. For instance, ``\textit{I look forward to collaborating with colleagues, exchanging experiences, and participating in dynamics that challenge us to think creatively and innovatively}'' \rebuttal{(S2P16)}, and \textit{``Practical, dynamic and lots of exchange of ideas}'' \rebuttal{(S2P21)}.

\subsection{RQ3: \rqthree} 
\label{sec:rq3}
This research question provides important results to compare with the expectations collected in Survey 2. The 23 Survey 3 respondents reported at the \textit{Demo Day} the main challenges they faced during the hackathon. They are:


\vspace{0.2cm}
\noindent
\textbf{Time Constraints}. One of the prominent challenges mentioned by participants was related to time constraints. Many felt that the limited timeframe posed a significant challenge in completing their tasks and staying organized. For instance, \rebuttal{S3P1} highlighted: ``\textit{I think the worst part was developing everything in 2 weeks}'', while \rebuttal{S3P2} stated, ``\textit{Our biggest villain was time. We had a little problem starting the project}.''

\vspace{0.2cm}
\noindent
\textbf{Group Dropouts and Lack of Commitment}. Some participants noted that the sudden departure of group members placed a strain on those who continued with the project, often leaving them to bear most of the work alone. To exemplify
, it is worth mentioning that the hackathon started with 50 participants and finished with 26. \rebuttal{S3P6} remarked: ``\textit{In fact, there were a lot of dropouts from my group, I was practically alone at the end}'', and \rebuttal{S3P22} shared, ``\textit{Because my group dropped out, getting through the Demo Day was a challenge}.'' 

\vspace{0.2cm}
\noindent
\textbf{Communication and Collaboration Issues}. Particularly when involving participants from different time zones or when dealing with disrespectful attitudes within the group. Comments such as ``\textit{Time zone differences and communication difficulties}'' \rebuttal{(S3P5)}, and ``\textit{I had several crises related to the lack of accessibility for autistic people... and personal disrespect} [among the group]'' \rebuttal{(S3P10)} 
pointed to the strain such issues placed on group dynamics.

\vspace{0.2cm}
Despite these difficulties, participants mentioned some factors that motivated them to finish the hackathon journey: 

\vspace{0.2cm}
\noindent
\textbf{Commitment and Group Support.} Team support played a critical role in sustaining their efforts. The motivation derived from supportive teammates was evident in responses such as: ``\textit{The team was essential to keep the spirits up}'' \rebuttal{(S3P18)}, and ``\textit{Show the young people in the group that we can do a good job even with little time and that we can deliver something of quality}'' \rebuttal{(S3P16)}.

\vspace{0.2cm}
\noindent
\textbf{Desire to Deliver the Final Project.} Working to deliver something at the end was also an important motivator. Participants expressed their determination with statements like ``\textit{The hope of delivering the project in the end}'' \rebuttal{(S3P21)}, showcasing their perseverance even amidst challenges. 

\vspace{0.2cm}
\noindent
\textbf{The Learning During the Journey.} The learning experience gained during the hackathon journey was another positive factor that kept participants engaged. \rebuttal{S3P2} reflected: ``\textit{The challenge of the new. I didn't know anything about AI. That was really wonderful}'', illustrating the excitement and value they found in learning something new.

\vspace{0.2cm}
\noindent
\textbf{Hiring and Recognition Expectations.} Lastly, expectations of hiring and recognition motivated participants to continue their efforts. Some viewed the hackathon as a pathway to potential career opportunities and validation of their work. This is captured in statements such as ``\textit{Learning and the scope of the project as well as the possibility of gaining recognition in a hackathon on such a pertinent and interesting subject}'' \rebuttal{(S3P7)}, and ``\textit{Maybe be selected to work at the company}'' \rebuttal{(S3P3)}, which pointed to their hopes for professional advancement and acknowledgment.

\subsection{RQ4: \rqfour} 

 \label{sec:rq4}
\begin{figure}[h!]
    \centering
    \includegraphics[width=1\linewidth,alt={The image is a bar chart comparing the ratings of three different aspects: Presentation, Topics chosen, and Facilitators. The ratings are divided into three levels: Could be better, Good, and Excellent. The y-axis represents the number of people, ranging from 0 to 30. Each rating category is represented by different colors: light peach for Could be better, light blue for Good, and dark blue for Excellent. The results shown are: Presentation: 0 ratings for Could be better, 4 ratings for Good, and 23 ratings for Excellent. Topics chosen: 0 ratings for Could be better, 6 ratings for Good, and 21 ratings for Excellent. Facilitators: 0 ratings for Could be better, 6 ratings for Good, and 21 ratings for Excellent.}]{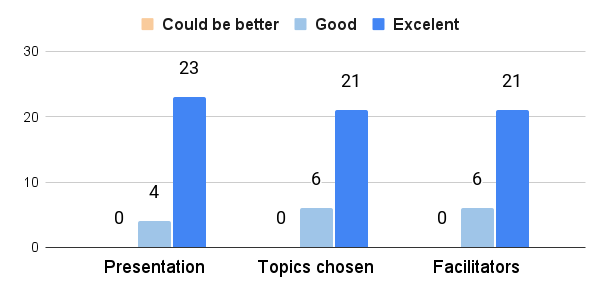}
    \caption{``\textit{Knowledge pills}'' experience evaluation}
    \label{fig:knowledgepills}
\end{figure}

We present in this section an overview of the factors that influenced successful collaboration, according to participants' perceptions. 
The following results highlight different perspectives of the participants' experiences:

\vspace{0.2cm}
\noindent
\textbf{About the ``\textit{Knowledge Pills}''.} 
Presentation, topics chosen, and facilitators' performance were mostly evaluated as excellent (respectively 23, 21, and 21 from 27 answers), with the remaining indicated as good (see Figure~\ref{fig:knowledgepills}). Among the 15 responses received, 10 praised the content, material, and topics presented. Improvements were suggested in making the material and tools available in advance (3), double-checking tools' accessibility (1), and reducing the number of pills to leave more time available to deliver the project (1).

\noindent
Of the 16 valid answers regarding the facilitators, 14 highlighted positive aspects, such as their knowledge of the topic presented, meeting dynamics, and capacity to answer the participants' doubts. Some positive feedback was: \textit{``Everyone seemed to be experts in their fields, they handled the meetings very well and answered questions effectively. Seeing} [name here] \textit{whose autism page I've been following for years, in the role of a respected instructor and professional in the field was especially important to me, as I'm facing challenges in the job market due to being autistic''} \rebuttal{(S3P10)}, and ``\textit{The way that content was explained was extremely clear and accessible. It really helped with understanding}'' \rebuttal{(S3P20)}.

\begin{figure}[h!]
    \centering
    \includegraphics[width=1\linewidth,alt={The image is a bar chart evaluating three aspects of an event: Online meeting tool, Accessibility support, and Overall event organization. Each aspect is rated in three categories: Could be better, Good, and Excellent. The bars are color-coded with peach for Could be better, light blue for Good, and dark blue for Excellent. The results are: Online meeting tool: 0 people rated it as Could be better, 4 people rated it as Good, and 23 people rated it as Excellent. Accessibility support: 1 person rated it as Could be better, 1 person rated it as Good, and 25 people rated it as Excellent. Overall event organization: 2 people rated it as Could be better, 6 people rated it as Good, and 19 people rated it as Excellent.}]{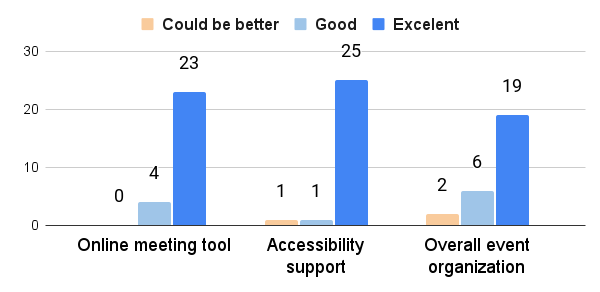}
    \caption{Event management evaluation}
    \label{fig:eventorganization}
\end{figure}

\vspace{0.2cm}
\noindent
\textbf{About Event Management.} We asked the participants to evaluate aspects connected to the event organization, such as accessibility promotion and the company's hackathon conduction. 
The online meeting tool was evaluated as excellent by 23 people and as good by four. Despite representing the better-evaluated criteria (25 responses as excellent), accessibility support also received one good, and one could be better feedback. Finally, the overall event organization was considered excellent by 19 and good by two participants, while two participants considered it could be better (see Figure~\ref{fig:eventorganization}).

From the open responses, 18 did not fill this field, and seven were evaluated positively: ``\textit{The use of Google Meet was really good due to its practicality. The schedule was equally good, and the organization was perfect}'' \rebuttal{(S3P18)}. Two people complained about accessibility issues (neurodivergent requirements not accomplished by the group and a sign language translator that could be better), and the other two complained about the technical mentor's absence. They stated the company should have worked harder to overcome these difficulties.  \rebuttal{S3P2} said: ``\textit{Nothing to say about accessibility, but regarding organization, yes. For example, the mentor assigned to our group simply didn't show up}''.
 
\vspace{0.2cm}
\noindent
\textbf{About Technical Mentors.} Both closed questions on this subject -- mentors' performance and support on the project's activities -- received the same evaluation: 23 excellent, two good, and two could be better. Reading the open-question answers, one can notice the negative evaluations were concentrated in one isolated case where the mentor's absence fostered inequities inside the group. Among participants, 14 people evaluated the role of the mentor positively, and many mentioned them by name: ``\textit{Our mentor was great. He was always available whenever we need help. Besides that, he shared a lot of experience through advice and tips}'' \rebuttal{(S3P19)}, and ``\textit{I just have praise for [Name Here], he was outstanding, offering advice not just for the project, but also for my career and life}'' \rebuttal{(S3P15)}.

\rebuttal{To improve the understanding of the experience, we collected participant's perceptions based on the challenges that \textit{Survey 2} respondents pointed out at S2Q4 (see Table ~\ref{tbl:survey2-questions}) as faced in previous collaborative initiatives.}

 
These factors are: \textit{Ineffective communication} that caused confusion, misalignment, and missed collaboration opportunities, wasting time and effort. \textit{Lack of empathy} that led to individual needs being ignored, especially regarding accessibility for neurodiverse participants.
\textit{Poor task distribution and organization} that delayed progress, particularly affecting beginners who struggled to follow group dynamics.
\textit{Lack of openness to questions and contributions} that made participants hesitant to engage, allowing only a few voices to dominate.
\textit{Lack of other participant's interest} that slowed the group, frustrating those committed to the task.
\textit{Other:} such as excessive competition, unclear goals, and lack of structured learning support that hindered collaboration and progress.

Now, we will present the comparative results of the participants' perceptions on key criteria \rebuttal{based on four of these challenges: \textit{Ineffective communication}, \textit{Lack of empathy}, \textit{Poor task distribution and organization} and \textit{Lack of openness to questions and contributions}. For each one, we present the participant's evaluation under }two perspectives: from the company and among their group.

\begin{figure}[h!]
    \centering
    \includegraphics[width=1\linewidth,alt={The image is a bar chart that compares responses to a survey question about satisfaction levels on communication, divided into two categories: With the company members and Among the members of my group. For the 1 star category, there were no responses for With the company members, whereas Among the members of my group recorded 5 responses. In the 2 stars category, With the company members received 4 responses, and Among the members of my group also received 5 responses. For the 3 stars category, With the company members had 22 responses, while Among the members of my group recorded 16 responses. Lastly, for the did not answer option, both categories, With the company members and Among the members of my group, had 1 response each.}]{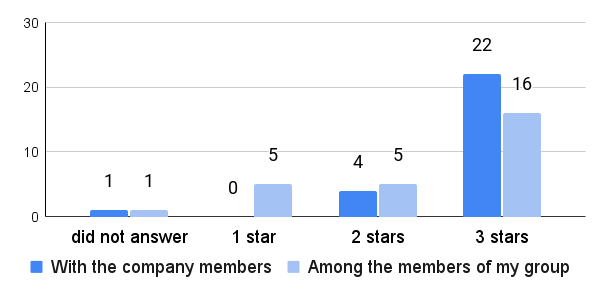}
    \caption{Communication aspects.}
    \label{fig:communication}
\end{figure}

\vspace{0.2cm}
\noindent
\textbf{Communication.} 22 respondents rated communication with the company three stars, four gave two stars, and one did not respond. For group communication, 16 gave three stars, five rated it one star, five gave two stars, and one did not respond (see Figure~\ref{fig:communication}). This suggests communication with the company was rated more positively than within groups.

\begin{figure}[h]
    \centering
    \includegraphics[width=1\linewidth,alt={The image is a bar chart that compares responses to a survey question about satisfaction levels on empathy, divided into two categories: With the company members and Among the members of my group. For the 1 star category, there were no responses for With the company members, whereas Among the members of my group recorded 3 responses. In the 2 stars category, both categories received 3 responses. For the 3 stars category, With the company members had 23 responses, while Among the members of my group recorded 18 responses. Lastly, for the did not answer option, With the company members had 1 response and Among the members of my group, 3.}]{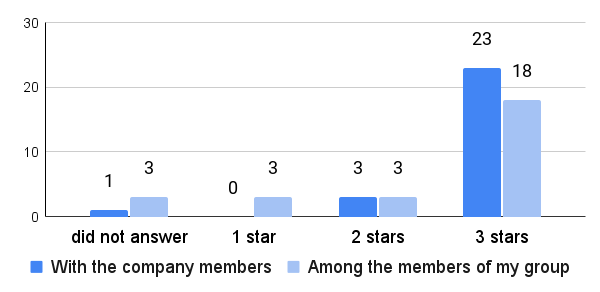}
    \caption{Empathy aspects.}
    \label{fig:empathy}
\end{figure}

\vspace{0.2cm}
\noindent
\textbf{Empathy.} One respondent did not answer about interactions with company members; three rated it one star, three gave two stars, and 23 gave three stars. For group interactions, three did not answer, three rated it one star, three gave two stars, and 18 gave three stars (see Figure~\ref{fig:empathy}). This suggests empathy was perceived more positively in interactions with company members than among participants.

\begin{figure}[h]
    \centering
    \includegraphics[width=1\linewidth,alt={The image is a bar chart that compares responses to a survey question about satisfaction levels on organization aspects, divided into two categories: With the company members and Among the members of my group. For the 1 star category, there was 1 response for With the company members, whereas Among the members of my group recorded 5 responses. In the 2 stars category, With the company members received 2 responses, and Among the members of my group received 4 responses. For the 3 stars category, With the company members had 23 responses, while Among the members of my group recorded 17 responses. Lastly, for the did not answer option, both categories had 1 response each.}]{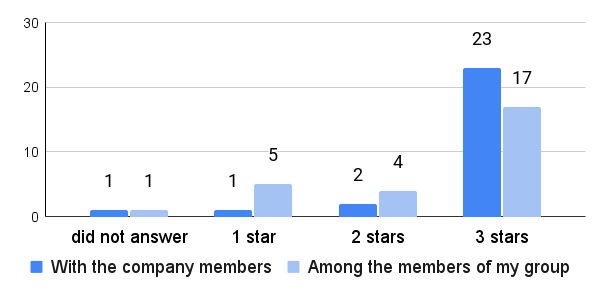}
    \caption{Organization aspects.}
    \label{fig:organization}
\end{figure}

\vspace{0.2cm}
\noindent
\textbf{Organization.} One respondent did not answer, one rated the company's organization with one star, two gave two stars, and 23 gave three stars. For group organization, there was one non-response, five rated it one star, four gave two stars, and 17 gave three stars (see Figure~\ref{fig:organization}). This suggests the company's organization was well-regarded, while group self-organization had a slightly less favorable perception.

\begin{figure}[h]
    \centering
    \includegraphics[width=1\linewidth,alt={The image is a bar chart that compares responses to a survey question about satisfaction levels on openess, divided into two categories: With the company members and Among the members of my group. For the 1 star category, there was no response for With the company members, whereas Among the members of my group recorded 2 responses. In the 2 stars category, both categories had 3 responses. For the 3 stars category, With the company members had 23 responses, while Among the members of my group recorded 20 responses. Lastly, for the did not answer option, with the company members had 1 response and Among the members of my group, 2.}]{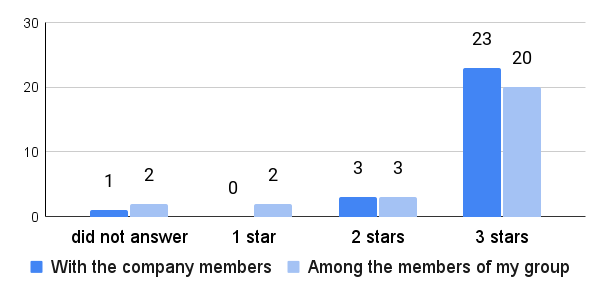}
    \caption{Openness aspects.}
    \label{fig:openness}
\end{figure}

\vspace{0.2cm}
\noindent
\textbf{Openness for Questions and Contributions.} 
One respondent did not answer, three rated it two stars, and 23 gave three stars for interactions with company members. For group interactions, two did not answer, two rated it one star, three gave two stars, and 20 gave three stars (see Figure~\ref{fig:openness}). The data shows a high level of openness in both contexts, but interactions with company members were rated more positively.

\vspace{0.2cm}
As a summarization of the satisfaction about the \textit{Affirmative Hackathon}, we noticed that 33\% of the \textit{Demo Day} participants considered it was \textit{as they expected} and 67\% considered it better than expected. As a positive balance, 92\% would recommend to a friend to participate in the next editions.

\section{Discussion} 

Hackathons are examples of time-bounded collaborative events where participants work together for a short period to achieve specific goals~\cite{FutureHackJeanette}. There are different types of hackathons (educational, civic, and corporate)~\cite{PeThan2019}. In this study, we focused on a corporate external initiative~\cite{9226320}. 

After analyzing the initial participants' profiles, we could create the \companyName's hackathon persona: a ``\textit{white cisgender heterosexual man, with a physical disability, identified as a beginner software developer with one or two years of experience in back-end Java technology}''. Although we cannot make statistical claims about our results, it is interesting to notice that, even in the PWD community, the profile is one of a white cisgender heterosexual man. While this result might appear to be due to selection bias, \companyName used different approaches to reach PWD participants, such as sharing the invitation on social networks and also reaching out to non-governmental organizations to advertise the event (more on Section~\ref{sec:preparation}). 
Perhaps this finding is not surprising since most computer science and software engineering enthusiasts fit this profile. For instance, in a 2017 survey with 5,500 GitHub users, 90\% of them self-identified as man~\cite{githubsurvey}. In any case, given that the technology sector has emerged as a field to employ PWD, this result deserves attention and reinforces previous studies'~\cite{intersec, intersec2} recommendations in deepening the PWD inclusion considering intersectionality in diversity. \rebuttal{Overall, one of the lessons learned is that companies interested in organizing affirmative hackathons should adopt different strategies for attracting a diverse set of PWD.}

Our second research question (Section ~\ref{sec:rq2}) focused on the participants' motivations and expectations for the hackathon. Most of the motivations we report are similar to non-PWD's, such as: improving technical knowledge and networking opportunities~\cite{AlmeidaEtal2021, Perin2024, PeThan2019}. However, we identified a new motivation in our public: the possibility to engage with other PWD. In this case, it is worth reflecting on the importance of representation to minoritized groups, which has already been the object of previous research in different contexts~\cite{TransHackathon, kievHackathons}. Furthermore, besides the interaction among each other, having PWD on the organizing team can create a more empathetic and inclusive environment, as they bring firsthand understanding and insights into accessibility and inclusion challenges and act as an inspiration for participants. \rebuttal{In other words, having PWD on the organizing team is an important lesson we learned and should be considered in similar initiatives.}

Despite reporting the participation in the hackathon as a \emph{motivation} to find a job, the 
participants did not mention any \emph{expectation} related to it, which may suggest a low level of self-confidence or uncertainty regarding their chances of success in the job market, likely due to past experiences or perceived inclusion challenges in the tech industry. However, expectations and motivation are closely related concepts, and participants may have inadvertently mixed these notions. We identified mentions about hoping for hiring and recognition when analyzing what kept them motivated to finish the challenge, even considering the difficulties they faced during the journey. 
\rebuttal{We plan to clearly differentiate these concepts in our future research on this topic, i.e., this is an important aspect to be followed by other companies.}

The participant's dropouts called our attention since from the ten initial groups, four dropped out of the \textit{Demo Day}. Further research should investigate to understand the reasons for these dropouts (e.g., whether they are connected to accessibility or any other factor related to full-remote hackathons).

Communication and collaboration raised issues linked to PWD-specific matters also were perceived as a surprise since we expected that empathy among PWD could facilitate these aspects. Some insights can be obtained by analyzing key factors (Section ~\ref{sec:rq4}), where all the collected feedback about communication, empathy, organization, and openness were better evaluated when considering the company's approach, over the group colleagues. 

\rebuttal{Thus, companies interested in organizing affirmative hackathons should be prepared to manage group dynamics difficulties, and apply measures to reduce or avoid, dropouts.}

Our final reflection about this experience relies on the relevance of corporate affirmative hackathons as valid instruments to encourage and support PWD with knowledge, network, and job opportunities, collaborating for reducing inequities in the work market and expanding awareness about accessibility and effective inclusion~\cite{rocha2024effective} in the software development industry.

\section{Limitations}


\rebuttal{Our work limited the participant's profile on full-stack development knowledge, which could have restricted the applicant pool to those with broader skill sets. Consequently, more specialized developers may have been excluded, reducing the group’s overall diversity.}

In addition, 
collecting only survey data to capture participants' experiences and perspectives. While surveys provide important insights, they may not fully convey the depth of participants' thoughts, motivations, or challenges~\cite{creswell2013research}. 
This could limit the nuanced understanding of the participants' experiences and perspectives that could emerge from a more interactive format. Furthermore, our survey did not use a motivation scale for time-bounded collaborative events~\cite{Perin2024}, because our goal was beyond PWD's motivation.

\rebuttal{Still, it is important to note that around 20 participants did not complete the hackathon. This dropout rate could incur bias, limiting the diversity of perspectives captured in the final dataset.} 

The lack of input from the hackathon organizers is another limitation of this work, i.e., we did not collect feedback from the event's coordinators, mentors, or other organizing team members. Their perspectives on the hackathon would offer an additional 
context, helping us to understand the effectiveness of the organizational strategies in place for the participants. 

This study did not examine the technical solutions created by participants during the hackathon. Reviewing these solutions would have provided information on participants’ problem-solving approaches. Since the data collection for Surveys 1 and 2 was anonymous, we could not link this ``assessment'' information with profile data.
Similarly, this study does not account for long-term outcomes in participants’ careers. We believe that a longitudinal study~\cite{creswell2013research} is an interesting research direction to pursue, which we leave for future work.

\rebuttal{Finally, our findings might not be transferred to other tech companies, particularly because each environment's unique organizational culture, resources, and constraints may lead to differing outcomes, limiting the generalizability of our results.}

\section{Conclusion}

This paper presents an exploration of an industry-led hackathon designed to empower software developers with disabilities (PWD). The hackathon, organized by \companyName, aimed to foster inclusion in the technology sector by providing a collaborative environment where participants could showcase their technical skills, connect with peers, and potentially join the company's talent pool. As direct results, ten participants were hired and are now working as developers in \companyName projects. Other 146 were added to the talent pool, being considered for further opportunities.
Through the use of three structured surveys, we gathered data on participant demographics, motivations, challenges faced, and overall experiences, with the goal of drawing lessons that could inform future events and industry practices.

Our analysis reveals key insights into the profile and experiences of hackathon participants. The majority identified as white, cisgender, heterosexual men, indicating a common demographic trend in technology groups, even within PWD-specific initiatives. Motivations centered around career growth, technical learning, and networking opportunities. 
Challenges encountered included time constraints, group member dropouts, and accessibility issues, highlighting the complex dynamics in team-based hackathons. Despite these obstacles, factors such as commitment, the desire to learn, and group members' support contributed to the hackathon's successful completion. The findings underscore the importance of addressing accessibility needs and fostering a collaborative, supportive environment to optimize participant experiences and outcomes, especially in a full-remote hackathon.


Future work should expand on these findings by incorporating qualitative interviews with participants and event organizers to provide deeper insights into their experiences and organizational strategies. Examining the technical solutions created during the hackathon could also offer a clearer picture of participants’ problem-solving capabilities and teamwork dynamics that resulted from the initiative. 


\section*{Acknowledments}
\label{sec:acknowledgments}
The authors thank CNPq (projects 400920/2019-0, 308623/2022-3, and 420406/2023-9) for supporting this research.


\bibliographystyle{IEEEtran}
\bibliography{bibliography.bib}

\end{document}